\documentclass[algorithms,article,accept,moreauthors,pdftex]{mdpi} 
\usepackage{amsfonts}
\usepackage{mathtools}
\usepackage{physics}
\usepackage{bm}
\usepackage[vlined,ruled]{algorithm2e}
\def\bra#1{\mathinner{\langle{#1}|}}
\def\ket#1{\mathinner{|{#1}\rangle}}
\def\braket#1{\mathinner{\langle{#1}\rangle}}

\usepackage{amsmath,amssymb}
\usepackage{ulem} 
\usepackage{latexsym}
\newcommand{\BH}[1]{\textcolor{black}{#1}}

\firstpage{1} 
\pubvolume{1}
\issuenum{1}
\articlenumber{0}
\pubyear{2020}
\copyrightyear{2020}
\history{Received: date; Accepted: date; Published: date}
\Title{Searching via nonlinear quantum walk on the 2D-grid}
\Author{Giuseppe Di Molfetta  $^{1,}$* and Basile Herzog $^{1,2}$}
\AuthorNames{Firstname Lastname, Firstname Lastname and Firstname Lastname}

\address{%
$^{1}$ \quad Aix-Marseille Université, CNRS, LIS, Marseille, France; e-mail@e-mail.com\\
$^{2}$ \quad Université de Lorraine, LPCT, Nancy, France; e-mail@e-mail.com}

\corres{giuseppe.dimolfetta@lis-lab.fr}

\abstract{We provide numerical evidence that the nonlinear searching algorithm introduced by Wong and Meyer \cite{meyer2013nonlinear}, rephrased in terms of quantum walks with effective nonlinear phase, can be extended to the finite 2-dimensional grid, keeping the same computational advantage with respect to the classical algorithms. For this purpose, we have considered the free lattice Hamiltonian, with linear dispersion relation introduced by Childs and Ge \cite{Childs_2014}. The numerical simulations showed that the walker finds the marked vertex in $O(N^{1/4} \log^{3/4} N) $ steps, with probability $O(1/\log N)$, for an overall complexity of $O(N^{1/4}\log^{5/4}N)$, using amplitude amplification. We also proved that there exists an optimal choice of the walker parameters to avoid that the time measurement precision affects the complexity searching time of the algorithm.}

\begin{document}
\section{Introduction}
Searching an element among an unstructured database of size $N$ takes $O(N)$ iterations, resulting in a linear complexity time. In 1996, L. Grover came up with a quantum algorithm that speeds up this  brute force $O(N)$ problem into a $O(\sqrt{N})$ problem \cite{grover1996fast}. The algorithm comes in many variants and has been rephrased in many ways, including  quantum walks \cite{childs2004spatial}. Quantum walks (QW) are essentially local unitary gates that drive the evolution of a particle on a graph \cite{venegas2012quantum}, and although they may appear defined in a discrete and in a continuous time setting, it has been recently shown that a new family of "plastic" QW unify and encompass both systems \cite{di2020quantum, manighalam2020continuous}. They have been used as a mathematical framework to express many quantum algorithms, e.g. \cite{ambainis2007quantum, childs2003exponential,childs2009universal}, but also many quantum simulation schemes e.g. \cite{hatifi2019quantum,di2016quantum}. In particular, it has been shown that many of these QW admit, as their continuum limit, the Dirac equation \cite{di2013quantum,arrighi2018dirac}, providing ‘quantum simulation schemes’, for the future quantum computers, to simulate all free spin-1/2 fermions. More interestingly, it has been recently proven by one of the authors, that the Grover algorithm is indeed a naturally occurring phenomenon, i.e. spontaneously implemented by some kind of particles in nature \cite{roget2020grover} over arbitrary surfaces with topological defects. From a theoretical perspective, a Grover search on a graph, rephrased in terms of QW, is an alternation of a diffusion step and an oracle step. The nodes of the graph represent elements of the configuration space of a problem, and whose edges represent the existence of a local transformation between two configurations. So far, the QW search has only been used to look for ‘marked nodes’, i.e. good configurations within the configuration space, as specified by an oracle. In \cite{roget2020grover}, it has been proved that instead of using them to look for ‘good’ solutions within the configuration space of a problem, we could use them to look for topological properties of the entire configuration space.   \\
The generalisation to an interacting multi-walkers scenario has already been explored in quantum algorithmics \cite{childs2013universal}, showing that such systems are capable of universal quantum computation. Moreover, there exists many physical systems which may be described by a nonlinear \textit{effective} equation, such as Bose-Einstein condensates (BEC)\cite{PhysRevA.88.032310, kevrekidis2007emergent}. Quite interestingly, the experimental setup proposed by Alberti and Windemberg in \cite{alberti2017quantum}, showed that a spinor BEC can simulate, in quasi-momentum space, a 1D QW with weak non-linearities.  Notice that such results, does not aim to suggest a scheme for implementing nonlinear quantum mechanics, which we know is linear. However it paves the way to simulate, via a mean-field approach, one particle nonlinear dynamics, where nonlinearities comes from the weakly interacting BEC.
In the following, in order to avoid confusion, we will introduce nonlinearities in the very same spirit of \cite{Abrams_1998}. Abrams and Lloyd consider a hypothetical nonlinear quantum mechanics, \textit{i.e.} a nonlinear Schrödinger equation for closed single quantum mechanical systems. In particular they show that nonlinearity in quantum computation, could make quantum systems solve NP-complete problems in polynomial time. 
Nonlinearity in Quantum Walks have been considered in several recent studies and may appear either under the form of nonlinear phases, (e.g. in Kerr medium \cite{molfetta2015nonlinear} and \cite{Navarrete_Benlloch_2007}) or via a feed-forward quantum diffusion operator \cite{shikano2014discrete}. In this manuscript we will give numerical and analytical evidence that nonlinearity leads to a clear computational advantage on the two dimensional grid with respect to the linear case, consistently with previous results on complete graph \cite{wong2015nonlinear}, paving the way to extend our nonlinear scheme to higher dimensional physical dimensional grids. \\
The manuscript is organized as follows : In Section \ref{sec:model}, we will introduce the QW in continuous time and discrete space; in Section \ref{sec:results} we will present our numerical simulations for the linear and nonlinear algorithm and in Section \ref{sec:anal} we will derive analytically the searching time. Finally, in Section \ref{sec:discussion}, we discuss the results and conclude. 

\section{Model}\label{sec:model}

\subsection{The linear algorithm}
Let us consider a quantum walk in continuous time, over a two dimensional grid of size $L \times L$, where $N=L^2$ is the number of vertices $v\in V =\{1,...,N\} $. The Hilbert space of the quantum walk is spanned by the set of vertices, such that:
\begin{equation}
    \mathcal{H} := span\{\ket{v} : v\in V\}.
\end{equation}
A general state of the walker $\ket{\psi(t)}$ is a vector, whose complex components $\psi_v = \braket{v}{\psi}$, evolve according the Schrödinger equation: 
\begin{equation}
    i\frac{d \psi_v(t)}{dt} = \sum_w H_{vw} \psi_w(t),
\end{equation}
where $H_{vw}$ are the coefficients of the Hamiltonian. We consider that $H$ is local, or in other terms that $H_{vw}\neq 0$ if and only if $v$ and $w$ are adjacent.

\begin{figure}[h!]
    \centering
    \includegraphics[width = 0.3\textwidth]{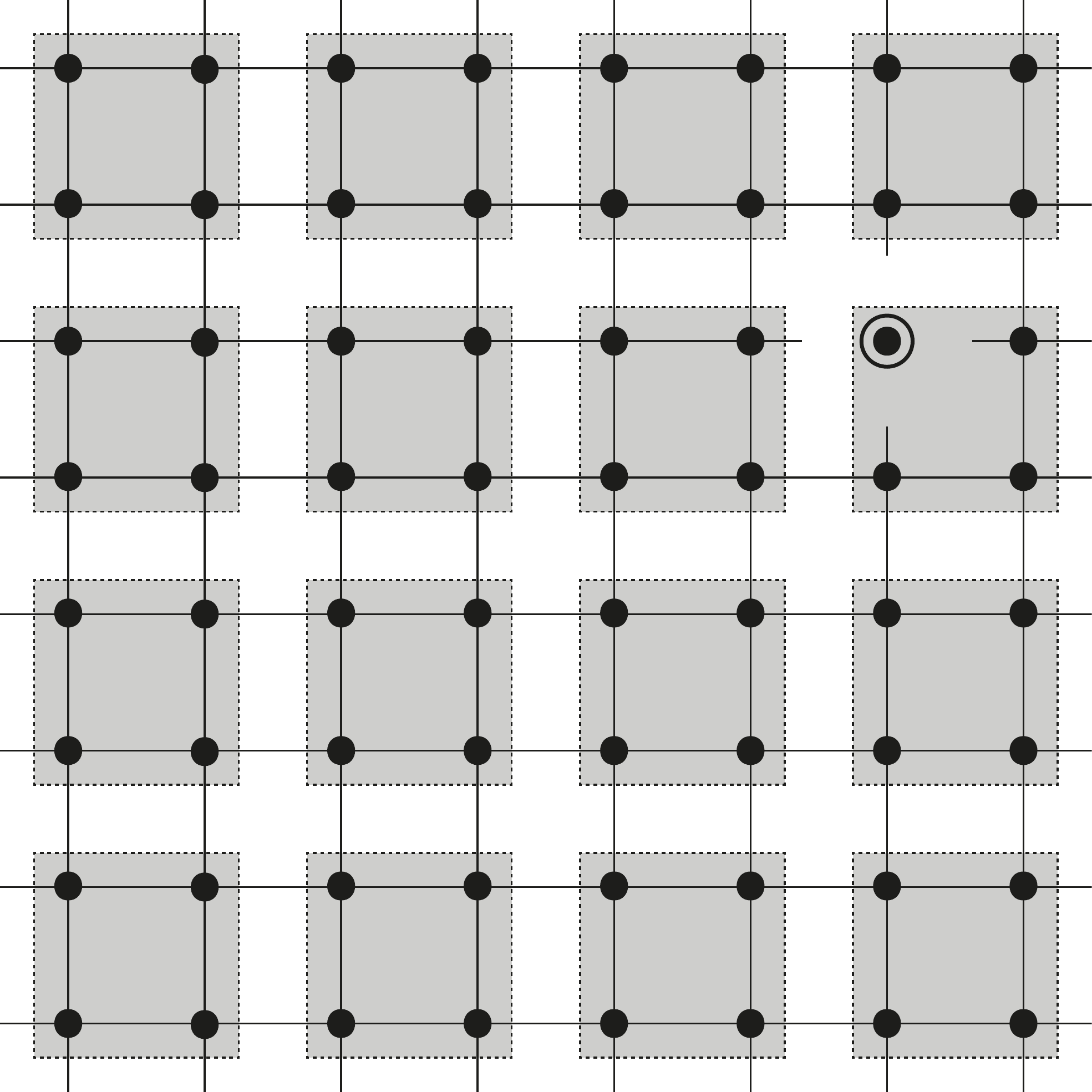}
    \caption{The $2d-$grid and its factorisation into cells (grey squares). The marked vertex (circled) is disconnected from its neighbors.}
    \label{grid}
\end{figure}

In this framework, the search problem corresponds to find a marked vertex $\ket{\bar v}$, given an initial state $\ket{\psi(t=0)}$, usually constructed as an uniform superposition over all vertices : 
\begin{equation}
    \ket{\psi(0)} = \frac{1}{\sqrt{N}} \sum_{v \in V} \ket{v}.
\end{equation}
The main idea is to let the quantum walker evolve on the grid for such a time that a measure of his state, on the computational basis, can be used to find the marked element with constant probability. Thus, we need to find the shortest time $T$, such that the probability $p(t)=|\bra{\bar v}e^{-i H t}\ket{\psi(0)}|^2$ is maximised. In the following, we refer to $T$, as the searching time and $\bar p = p(T)$, the success probability. In a continuous time setting, in order to search a single marked node, we let the walker evolve under the action of the following Hamiltonian:
\begin{equation}
H_L = H_0 + H_{oracle},
\label{eqn:search}
\end{equation}
where $H_0$ is the lattice free Hamiltonian of the walker and $H_{oracle}$ is the oracle Hamiltonian which is used to perturb the walker free Hamiltonian upon the marked vertex. In the original work by Childs and Goldstone \cite{Childs_2004}, it has been proven that, for a two dimensional lattice, the quadratic speed up with respect to the classical search algorithm is lost. Later, the same authors proved in \cite{childs2004dirac} that by introducing a second register, the coin state, as in the discrete time quantum walk, the quantum speedup $O(\sqrt{N}\log N)$ is achieved even for a two dimensional lattice. The reason has been investigated in \cite{childs2004spatial}, where again Childs and Goldston, inspired by Ambainis, Kempe and Rivosh \cite{ambainis2004coins}, pointed out that the optimality loss in two spatial dimension for a continuous time quantum walk came from the quadratic dispersion relation of the free Hamiltonian. Since that, several other results confirmed this interpretation, as for example in \cite{foulger2014quantum}, where a coinless walker with linear dispersion relation over a honeycomb lattice led to the same quantum speedup. In this paper, we will use the same model introduced by Childs and Ge \cite{Childs_2014}, where a linear dispersion law is achieved introducing periodic inhomogeneities into the lattice Hamiltonian : the idea is to group, periodically, multiple neighbouring vertices into cells, as in Fig.\ref{grid}. In order to compare this model with the coined quantum walk, similarly to the staggered fermion model \cite{susskind1977lattice}, we can notice that the coin states may be seen as embedded into the cells. The dimension of the coinspace is mimicked by the number of nodes grouped in each cell. Although this construct may suggest that now the searching algorithm will look for a marked cell and no longer for a specific vertex, we will show that an opportune choice of the oracle $H_{oracle}$ allows us to look for any single vertex as possible marked item, keeping the advantage of the Hamiltonian over the cells. Let us now review more formally the model, as introduced in \cite{Childs_2014}. \\
Consider the lattice partitioned in $n := N/4$ cells,  as in Fig.\ref{grid}, with even $L$ and periodic boundary conditions. The free lattice Hamiltonian,
\begin{equation}
\begin{split}
    H_0\ket{\bm v} = (-1)^{v_x}(\ket{\bm v+ \bm{e_{x}}}-\ket{\bm v- \bm{e_{x}}})\\
    + (-1)^{v_x+v_y}(\ket{\bm v + \bm{e_{y}}}-\ket{\bm v- \bm{e_{y}}})
    \end{split}
\label{eqn:freeHam}
\end{equation}
with $\bm{e_{i}}$ the unit vector in the $i$-direction, is translation invariant of length $2$, and, due to the periodic conditions the above Hamiltonian is well defined on the border of the lattice. More formally :
\begin{equation}
    [H_0, T_i] = 0 \hspace{0.5cm} T_i\ket{\bm v} = \ket{\bm v + 2 \bm{e_{i}}}.
\end{equation}
Because, the 2D-grid is now factorised into cells, each composed by four vertices, it is convenient to introduce a new coordinates system. The cell will be labeled by $\bm r$, such that $\bm r \in [l]^2$ with $l = L/2$, and each of the four internal vertices will be labeled by $\bm{\sigma} \in \mathbb{Z}_2^2$. In particular, we can write the following transformations, which map the vertices coordinates into cell ones :
\begin{align}
    r_i &= \left \lfloor{\frac{v_i}{2}}\right \rfloor \BH{,} \\
    \sigma_i &= v_i - 2r_i
\end{align}
where $\left \lfloor{.}\right \rfloor$ denotes the floor function. Now, the Hilbert space is spanned by $\ket{\bm r, \bm \sigma}$ and the Hamiltonian reads:
\begin{equation}
\begin{split}
    H_0\ket{\bm r, \bm{\sigma} } = \sum_{i=x,y} (-1)^{s_i(\bm \sigma)} (\ket{\bm r+\sigma_i e_i, \bm \sigma+e_i} -\\ \ket{\bm r-\bar \sigma_i e_i,\bm \sigma+e_i})
    \end{split}
\label{eqn:freeHamNC}
\end{equation}
where $s_x(\bm\sigma) = \sigma_x$, $s_y(\bm\sigma) = \sigma_x + \sigma_y$ and $\bar \sigma_i = 1- \sigma_i$.\\

In this new setting, the searched vertex is :
\begin{equation}
    \ket{\bm{\bar{v}}} =  \ket{ \bm w, \bm{\alpha}} 
\end{equation}
with $\bm w \in [l]^2$ and $\bm{\alpha} \in \mathbb{Z}_2^2$. \\

We now define the oracle Hamiltonian, previously introduced in \cite{Childs_2014}, by : 
\begin{equation}
    H_{oracle} = -\ket{\bm w,\bm{\alpha}}\bra{\bm w,\bm{\alpha}}H_0 - H_0 \ket{\bm w,\bm{\alpha}}\bra{\bm w,\bm{\alpha}}
\label{eqn:oracle}
\end{equation}

Notice that the above oracle Hamiltonian, $H_{oracle}$, added to $H_0$, has the effect of disconnecting the marked vertex from its neighborhood, as in Fig. \ref{grid}. Indeed, by the definition, for any $\bm v$, we have $\bra{\bm v} H_0 \ket{\bm v} = \bra{\bm w,\bm{\alpha}} H_0 \ket{\bm w,\bm{\alpha}} = 0$, and it follows that 
\begin{eqnarray}\label{eqn:disconnect}
H_{L}\ket{\bm w,\bm{\alpha}}&=& (H_{oracle}+H_0)\ket{\bm w,\bm{\alpha}}= \\ \nonumber
-\dyad{\bm w,\bm{\alpha}} &H_0& \ket{\bm w,\bm{\alpha}} -H_0 \ket{\bm w,\bm{\alpha}} + H_0 \ket{\bm w,\bm{\alpha}} =0.
\end{eqnarray}
As consequence, one can not reach the neighbors of $\ket{\bm w,\bm{\alpha}}$ from itself. Moreover, one can not reach $\ket{\bm w,\bm{\alpha}}$ from its neighbors neither, because the total Hamiltonian $H_{L}$ is real and then it has to be symmetric. \\
Note that this oracle Hamiltonian is different from the one usually used, for example in \cite{Childs_2004}, $H_{oracle} \propto \dyad{\bm w, \bm \alpha}$. Indeed, it has been proven in \cite{oracle2, oracle, Childs_2014}, that this redefinition is justified by the $2^2$ cell-partition of the lattice and the consequent linear behavior of the dispersion relation near the ground state of $H_0$.
Moreover, the disconnection of the marked vertex from the rest of the lattice and the on-site potentials at the neighbours of the marked item, is reminiscent of the natural occurring searching algorithm studied in \cite{roget2020grover}, in the discrete time setting.\\

Now, following Eq.\eqref{eqn:disconnect}, looking at the marked vertex $\ket{\bm w,\bm{\alpha}}$, coincides to find a neighbour of $\ket{\bm w,\bm{\alpha}}$ with constant probability or with a sufficiently large probability that can be amplified with reasonable computational overhead. In other terms, we have to search the following state:
\begin{equation}
    \ket{\Gamma} = \dfrac{1}{2}H_0 \ket{\bm w,\bm{\alpha}}. 
\label{eqn:gamma}
\end{equation}

Childs and Ge \cite{Childs_2014} proved that the overlap of $\ket{\Gamma}$ with the evolved state $e^{-i H T}\ket{s}$ is $\Omega(1/\sqrt{\log(N)})$, with $T = O(\sqrt{N \log(N)})$, where the initial state $\ket{s}$ reads :
\begin{equation}
    s := \sqrt{\frac{4}{N}} \sum_{\bm r} \ket{\bm r,\bm \alpha}.
    \label{eqn:s}
\end{equation}
Let us shortly discuss this result, redirecting the interested reader to the detailed calculations in \cite{Childs_2014}.
In order to analyse the algorithm, we have to compute the overlap $|\bra{\Gamma}e^{-i H T}\ket{s}|$ and the searching time $T$ for which it is maximised. By analysing the spectrum of the Hamiltonian $H$, we can prove that $H$ has two reals eigenvalues:
\begin{equation}
    E_{\pm} \approx \pm \sqrt{\frac{16 \pi}{N\log N}},
\end{equation}
with the corresponding eigenstates $\ket{\psi_{\pm}}$:
\begin{equation}
    \ket{\psi_{\pm}} = \sqrt{R_\pm} (H_0 - \mathbb{I}_2 E_{\pm})^{-1}\ket{\bm w,\bm{\alpha}}
\end{equation}
where $\sqrt{R_\pm} := \bra{\bm w,\bm{\alpha}}H_0\ket{\psi_\pm} \neq 0$ and $\mathbb{I}_2$ is the Identity matrix of dimensions $2$. Now, taking the overlap of each of the eigenstates, with the initial state we get :
\begin{equation}
    \bra{\psi_\pm}\ket{s} = -\frac{\sqrt{R_{\pm}}}{E_\pm\sqrt{N/4}} \approx \mp \frac{1}{\sqrt{2}}
\end{equation}
where the last approximation holds for large $n$ and has been proven rigorously in \cite{Childs_2014}. From the above equation, we derive the approximated expression for the initial state:
\begin{equation}
\ket{s} \approx \frac{1}{\sqrt{2}} \left( \ket{\psi_-}-\ket{\psi_+} \right)
\label{eq:s}
\end{equation}
and letting the system evolve for $T = \frac{\pi}{2 |E_\pm|}$ : 
\begin{equation}
    e^{- i H T} \ket{s} \approx \frac{1}{\sqrt{2}} \left( \ket{\psi_-}+\ket{\psi_+} \right).
    \label{eq:evolveds}
\end{equation}
Finally the overlap 
\begin{equation}
|\bra{\Gamma} e^{- i H t} \ket{s}|^2 \approx \frac{\pi}{\log N} \sin^2(2|E_\pm| t),
\label{eqn:overlap}
\end{equation}
shows that the system will be sufficiently close to the state $\ket{\Gamma}$, in a time $T= \pi/2|E_\pm| =  O(\sqrt{N \log N})$. In other terms, the Eq. \eqref{eqn:overlap} shows us that the system evolves in the two dimensional space spanned by $\ket{s}$ and $\ket{\Gamma}$, and after a searching time $T = \frac{\pi}{2 |E_\pm|}$, the unitary $e^{- i H T}$ rotates the state from $\ket{s}$ to $\ket{\Gamma}$.
\\
We can now summarize the above in the following algorithm :

\begin{algorithm}[H]
\SetAlgoLined
 Initialization : \\
  \hspace{0.5cm} $\ket{\psi}=\ket{s}$\;
  \hspace{0.5cm} $t=0$\; 
 \While{$t < T = O(\sqrt{N\log N})$}{
    \hspace{0.5cm} $\ket{\psi} = \exp(-i H_{L}\Delta t) \ket{\psi}$\;
    \hspace{0.5cm} $t=t+\Delta t$\;
 }
 Measure $\ket{\psi}$
 \caption{Searching algorithm for the $2d$-grid}
  \label{al:linear}
\end{algorithm}

In order to study numerically the algorithm, we proceed as follows: (i) Prepare, the initial state as in Eq. \eqref{eqn:s} ; (ii) Let the walker evolve with time; (iii) Quantify the searching time $T(N)$ before the walker reaches its success probability $\bar p(N)$ of being localized in a ball of radius 1 around the marked element. Then, estimate this success probability, at fixed $N$ ; (iv) Characterize $T(N)$ and $\bar p(N)$, i.e. the way the success probability and the searching time depend upon the total number grid points.

\subsection{Adding a nonlinearity}

Here we will consider a hypothetical nonlinear quantum mechanics in the same spirit of \cite{Abrams_1998}. However, let us notice that there are physical systems in which multiple interacting quantum particles, under particular conditions, can be effectively described as a single particle obeying to the nonlinear Schrödinger equation : 
\begin{equation}
\partial_t \psi(\bm r, t) = (H_0 - g |\psi(\bm r, t)|^2) \psi(\bm r, t), 
\label{eqn:NLSE}
\end{equation}
where $H_0$ is the free Hamiltonian of the single particle and $g$ is the coupling constant of the non linear systems. The nonlinear term corresponds to a Kerr law \cite{kohl2008optical, zhang2010new}, acting as a self-potential. One concrete example of this family of quantum systems is the Bose–Einstein condensate (BEC) \cite{einstein1924quantentheorie, bose1924plancks}, effectively described by Eq.\eqref{eqn:NLSE} in the limit of large particle number \cite{erdHos2010derivation}. In the recent years, BECs have been realised in several experimental setups \cite{cornell2002nobel, nikuni2000bose, o2020quantum} and some of them inspired experimental realisation of quantum walks \cite{PhysRevLett.124.050502, alberti2017quantum}. In the following we will consider a BEC with attractive interaction, i.e. $g>0$, in the context of quantum algorithms. Wang and Meyer have previously investigated such systems applied to spatial search \cite{wong2015nonlinear, meyer2014quantum} on complete graphs. Although they found a reasonable improvement in time complexity over the linear case, their algorithm is not directly reducible to the 2-dimensional grid, keeping the same quantum speedup respect to the classical case. In fact, Wong and Meyer used the same linear Hamiltonian introduced by Farhi and Gutmann \cite{farhi1998analog}, which is not optimal on the square grid, as rigorously proved in \cite{Childs_2004}. In the following, we will show that the same nonlinear spatial search introduced by Wang and Meyer can be optimal, even in 2D, considering the free Hamiltonian \eqref{eqn:freeHamNC}. \\
The basic idea is to modify Eq.\eqref{eqn:search}, including a self-potential of the kind :
\begin{equation}
    H_{NL} = - g\sum_{\bm r, \bm \sigma} \abs{\braket{\bm r, \bm \sigma}{\psi(t)}}^2 \dyad{\bm r, \bm \sigma}
\label{eqn:Hnl}
\end{equation}
where the coupling positive constant $g$ amplifies the accumulation of the probability at the marked state due to $H_{oracle}$, speeding up the search. The overall time-dependent Hamiltonian now reads:
\begin{equation}
    H = H_L + H_{NL}.
\end{equation}

Similarly to the linear algorithm, here we have to compute the success probability $p$, or equivalently we have to find a searching time for which the overlap between $\Gamma$ and the evolved state $\psi(t)$ is maximised. Before doing so, we must ensure that the non-linear contribution is as reasonable as possible. In other words, we wonder whether it is possible for the system to continue to evolve in the subspace generated by $\ket{\Gamma}$ and $\ket{s}$. We will see that, this is possible at the cost of rescaling the spectrum at fixed $N$. 

Let us look into the subspace spanned by the two orthogonal vectors $\{\ket{\Gamma},\ket{s}\}$, in which $\ket{\psi(t)}$  can be decomposed as follows :
\begin{equation}
   \ket{\psi(t)} =  a(t)\ket{\Gamma} + b(t)\ket{s},
\end{equation}
where now, the probability amplitudes are time-dependent. Moreover, $\Gamma$ and $s$ are eigenstates of $H_{NL}$, in fact :
\begin{equation}
\begin{split}
    H_{NL}\ket{s} &  = 
     -g\sum_{\bm r, \bm \sigma} \abs{\braket{\bm r,\bm \sigma}{s}}^2 \ket{\bm r, \bm \sigma} \braket{\bm r, \bm \sigma}{s} \\
  & = -g \sum_{\bm r, \bm \sigma} \frac{4}{N} \ket{\bm r, \bm \sigma} \sqrt{\frac{4}{N}} = -g\frac{4}{N} \ket{s}
   \end{split}
\end{equation}
and,
\begin{equation}
\begin{split}
    H_{NL}\ket{\Gamma} &= -g\sum_{\bm r, \bm \sigma} \abs{\braket{\bm r,\bm \sigma}{\Gamma}}^2 \ket{\bm r, \bm \sigma} \braket{\bm r, \bm \sigma}{\Gamma} \\
    &= -g \sum_{\bm r, \bm \sigma} \frac{1}{4} \ket{\bm r, \bm \sigma} \mel{\bm r, \bm \sigma} {\frac{H_0}{2}}{\bm w,\bm \sigma} = -g \frac{1}{4} \ket{\Gamma}. 
    \end{split}
\end{equation}

In this subspace, the nonlinear Schrödinger equation reads:
\begin{equation}
    \partial_t \begin{pmatrix}a \\ b \end{pmatrix} = - i H \begin{pmatrix}a \\ b\end{pmatrix}
\end{equation}
where we have dropped $t$ parameter from the amplitudes to lighten the notation.

The total Hamiltonian is represented by the following matrix:

\begin{equation}
H = \left(
\begin{array}{cc}
\dfrac{-g|a|^2}{4} & E \\
E & \dfrac{-4g|b|^2}{N} \\
\end{array}\right),
\end{equation}
where $E=|E_\pm|$.

Remark that, by adding $\dfrac{g|a|^2}{4}$ to the diagonal coefficients, we only introduce a global phase, which does not change the dynamics. Thus, the rescaled total Hamiltonian is :
\begin{equation}
H' =    \left(
\begin{array}{cc}
0 & E \\
E & g\delta \\
\end{array}\right),
\end{equation}
with 
\begin{equation}
    \delta(t) = \left(
    \frac{|a|^2}{4}-\frac{4|\beta|^2}{N}
    \right).
\label{eqn:delta}
\end{equation}

The eingenvalues of the above matrix are :
\begin{equation}
    E'_{\pm} = \frac{1}{2}g\delta \pm \frac{1}{2} \sqrt{(g\delta)^2+4E^2}
\label{ev}
\end{equation}
with eigenstates, respectively:
\begin{equation}
    \ket{\psi'_{\pm}} = (\frac{g\delta}{2E} \pm \frac{\sqrt{(g\delta)^2+4E^2}}{2E},1).
\label{EV}    
\end{equation}

Notice that, if the perturbation $g\delta$ is sufficiently small and we rescale the spectrum, we can recover the eigenstates of the form $(\pm 1, 1)$ and thus, force the system to oscillate between them, similarly to the linear case. Indeed, let us multiply the Hamiltonian $H_L$ by a term $(1+ c(N)g\delta)$, with $c(N)$ a real function of $N$. The eigenvalue $E$, now transforms as follows :
\begin{equation}
    E \rightarrow  E(1+  c(N) g\delta).
    \label{trick}
\end{equation}
In order to make the eigenstates \eqref{EV} approximately of the form $(\pm 1,1)$, we have to impose $2E(1+  c(N) g\delta) \gg g\delta$ or equivalently 
\begin{equation}
      c(N)  \gg \frac{1}{2E} - \frac{1}{g\delta}.
      \label{eq:low}
\end{equation}
Moreover, an upper bound of $c(N)$ can be obtained as follows : 
\begin{equation}
\delta(0) =  \frac{|a(0)|^2}{4}-\frac{4|\beta(0)|^2}{N} = -4/N
\end{equation}
 since $\ket{\psi(t=0)} = \ket{s}$ and $|\beta(t)|^2 = |\braket{s}{\psi(t)}|^2$. Thus,
\begin{equation}
    c(N) < c_{max} = \dfrac{N}{4g}.
\end{equation} 

Finally, keeping only the leading term $1/2E$ in Eq.\eqref{eq:low}, we can bound $c(N)$ as follows 
\begin{equation}
    \frac{N}{4g} > c(N) \gg \frac{1}{2}\sqrt{\frac{N \log N}{16 \pi}} = \frac{1}{2E}.
\end{equation}

Notice that, the above inequality implies that $g$ cannot scale faster than $\sqrt{N}$.\\

Finally, we can summarise the nonlinear algorithm as follows : 

\begin{algorithm}[H]
\SetAlgoLined
 Initialization : \\
  \hspace{0.5cm} $\ket{\psi}=\ket{s}$\;
  \hspace{0.5cm} $t=0$\; 
 \While{$t < T'$}{
    \hspace{0.5cm} $\ket{\psi} = \exp \left(-i \Delta t \left(\left(1 + gc\delta \right)
    H_L+H_{NL} \right)\right)\ket{\psi}$\;
    \hspace{0.5cm} $t=t+\Delta t$\;
 }
 Measure $\ket{\psi}$
 \caption{Searching with a nonlinear algorithm for the $2d$-grid}
 \label{al:Nlinear}
\end{algorithm}

Contrary to the linear algorithm \eqref{al:linear}, an exact analytical treatment to explicitly calculate $T'$ is difficult. In the following, first, we will provide strong numerical evidence that such a search scheme allows a clear temporal advantage over the linear case, deriving numerically $T'$ and second, we will give an approximate analytical proof for it. In conclusion we will discuss the time measurement precision and how it affects the choice of the parameters.

\section{Numerical results}\label{sec:results}

\begin{figure}[h!]
    \centering
    \includegraphics[width = 0.5\textwidth]{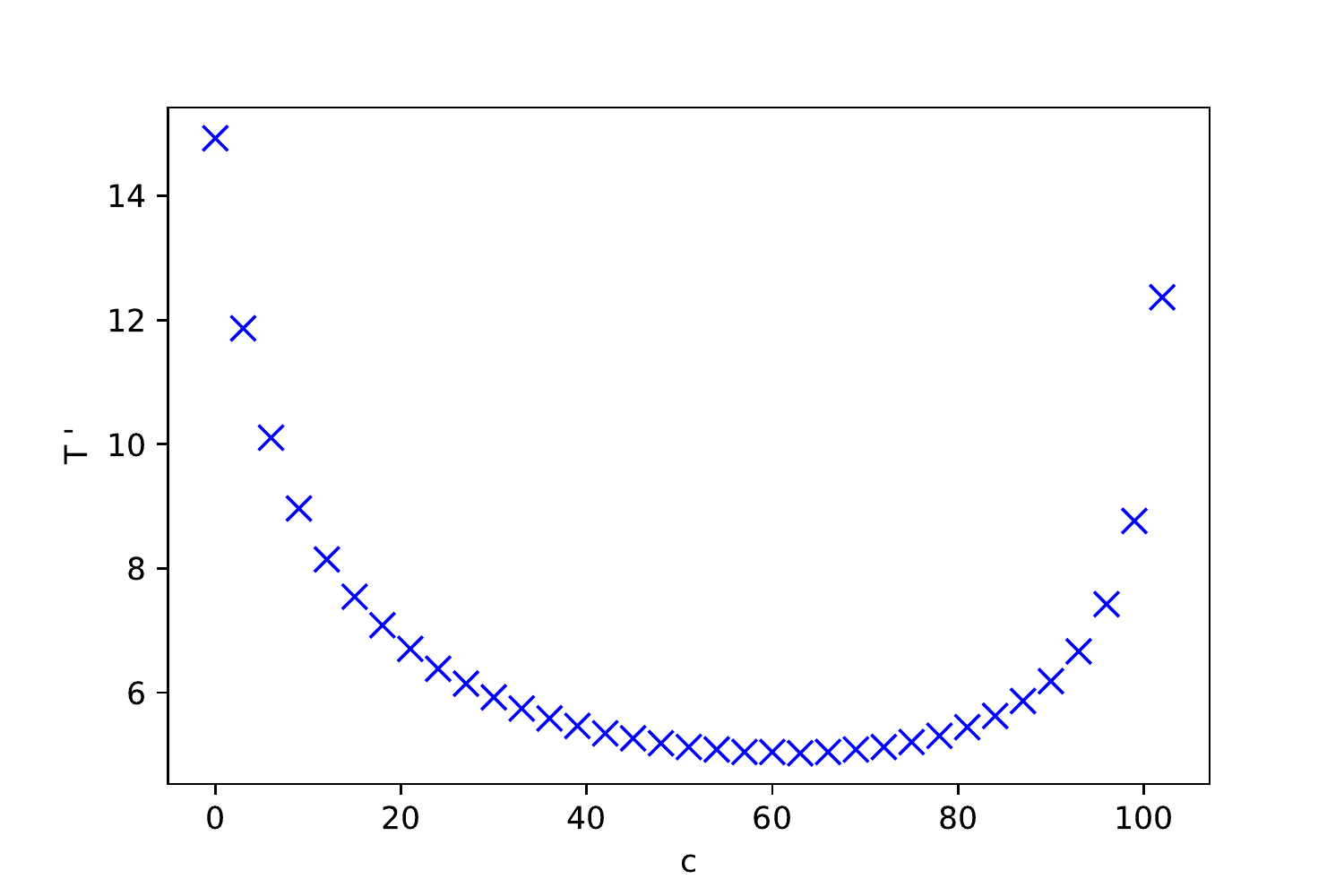}
    \caption{Searching time of the algorithm as a function of $c$, $N = 900$.}
    \label{c}
\end{figure}

\begin{figure}[h!]
    \centering
    \includegraphics[width = 0.5\textwidth]{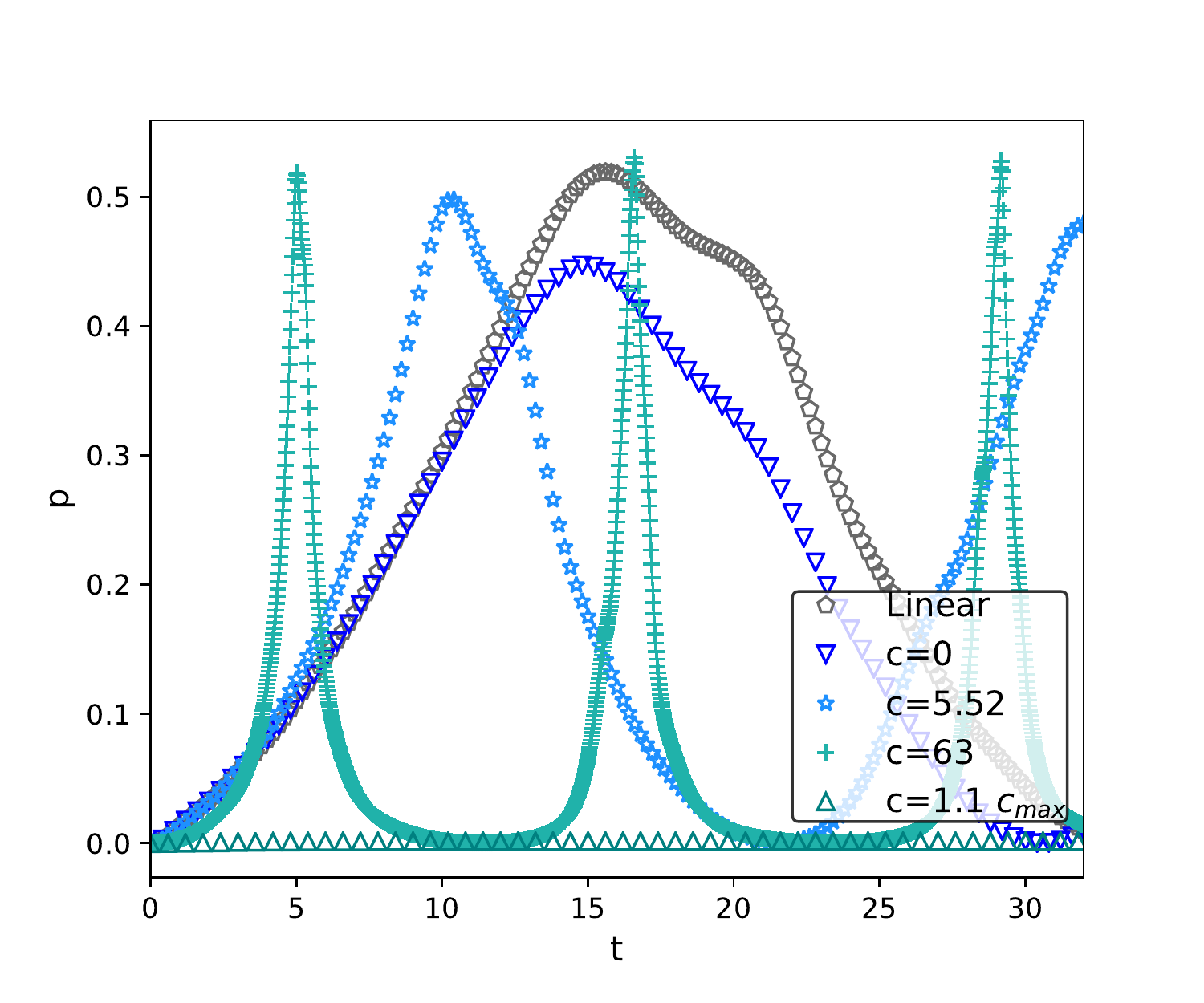}
    \caption{The probability $p(t)=|\bra{\Gamma}e^{-i H t}\ket{s}|^2$ for different values of $c$ - $N = 900$}
    \label{diff_c}
\end{figure}

Numerical simulations show in Fig.\ref{c} the searching time, $T'$ as a function of $c$ at fixed $N$ and, in Fig.\ref{diff_c}, the corresponding success probability for several values of $c$. In all cases, we have chosen $g=\log{N}/\pi$, which ensures that $g \delta \sim 1$. Notice that, for $c=0$ the searching time $T'$ is the same as for the linear algorithm, $T$, but the success probability is lower, because the initial eigenstate $\ket{s}$ is not rotated to the state $\ket{\Gamma}$, as in the linear case. In Fig.\ref{c}, we can also remark that the probability peak becomes narrower, the more $c$ increases, which would require more attempts to estimate the maximum of the probability curve, making the algorithm sub-optimal \cite{wong2015nonlinear}, as we discuss later. These considerations suggest to choose a value for $c$ which, while ensuring the good oscillatory behaviour of the system, leaves the peaks wide enough. Looking at Fig.\ref{diff_c}, a reasonable choice is  
\begin{equation}
c = 1/E = \sqrt{\frac{N \log N}{16 \pi}},
\end{equation}
(whose value is 5.52 for $N=900$ in Fig. \ref{diff_c}).
For the following, we will keep the above choice for our analysis.

\begin{figure}[h!]
    \centering
    \includegraphics[width = 0.5\textwidth]{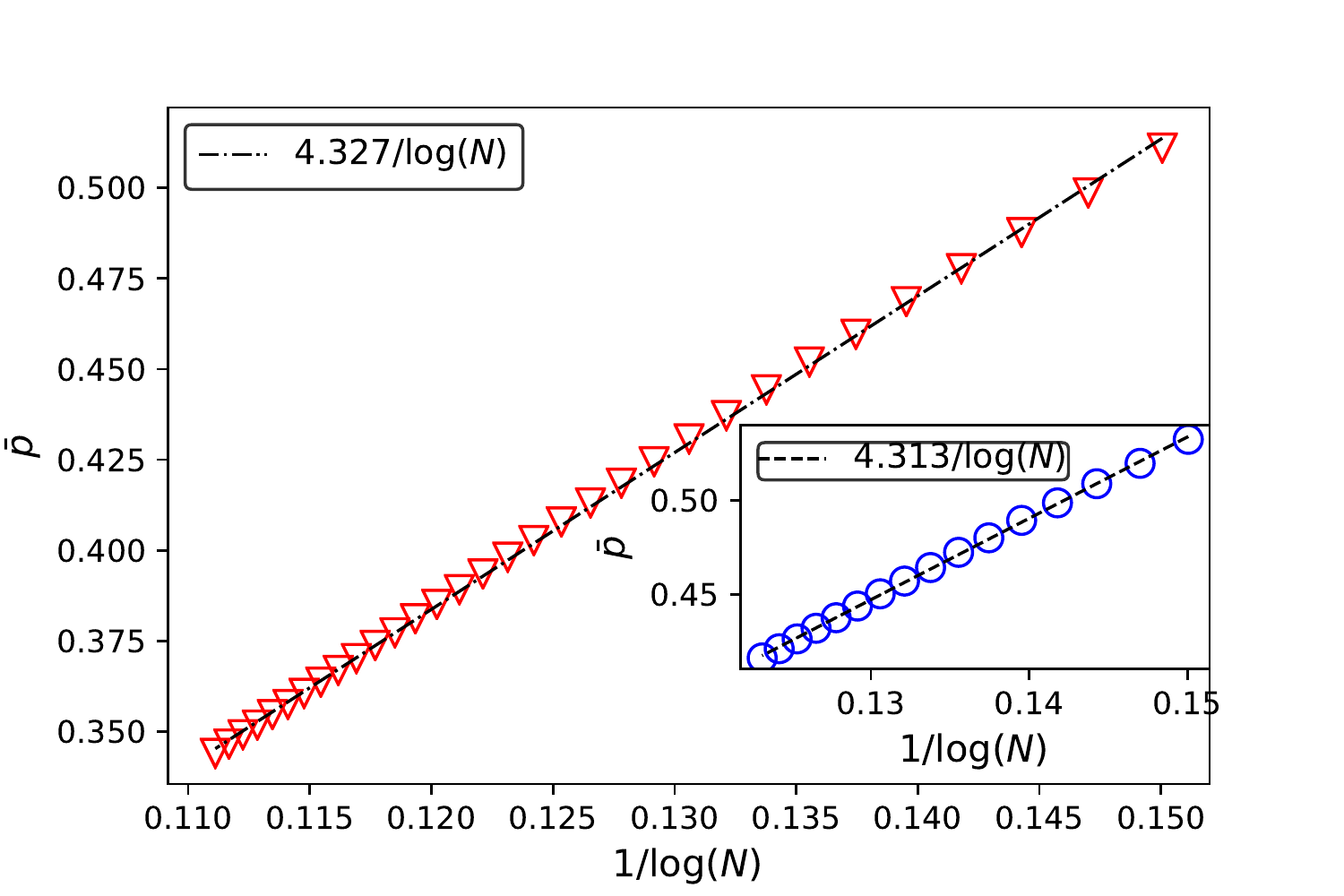}
    \caption{Success probability as a function of $N$ for the nonlinear and linear algorithm (inset)}
    \label{pofN}
\end{figure}

\begin{figure}[h!]
    \centering
    \includegraphics[width = 0.5\textwidth]{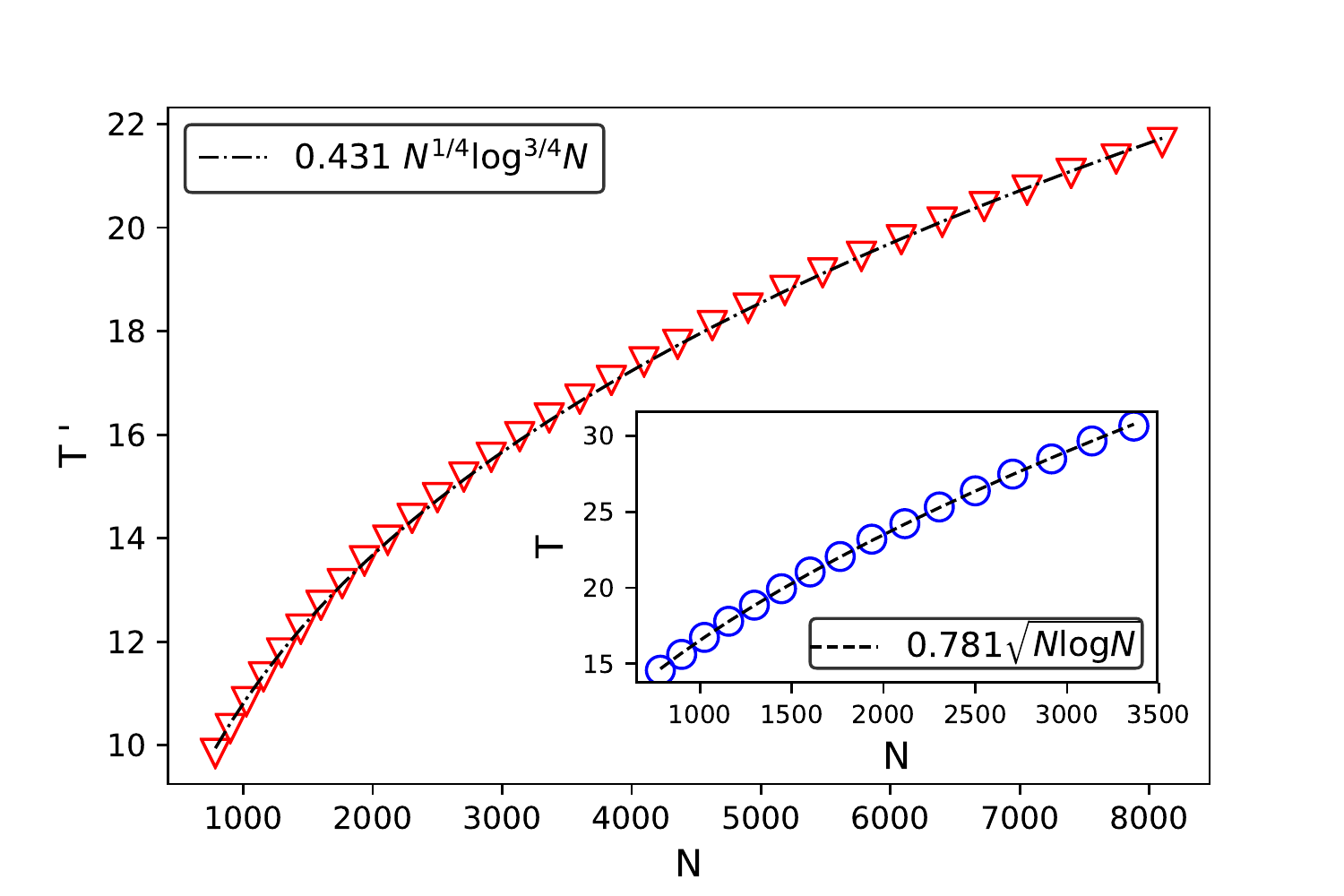}
    \caption{Searching times in the nonlinear and the linear algorithm (inset).}
    \label{tofN}
\end{figure}

Now we run both algorithms, the linear and the nonlinear one and we derive from both, a numerical characterization for the searching times $T(N)$ (linear), $T'(N)$ (nonlinear), and the success probabilities $\bar p(N)$. The first peak probability in the linear and in the nonlinear case, behaves as $1/\log(N)$, as it is shown in Fig.\ref{pofN}, approximately with the same pre-factor $\sim 4.3$.
The Fig.\ref{tofN} shows the searching times, $T$, and $T'$. As expected, in the linear model, we recover $T = O(\sqrt{N \log N})$. The nonlinear case, instead, shows a significant advantage respect to the linear case, with 
\begin{equation}
    T' = O(N^{1/4} \log^{3/4} N),
\end{equation}
which yields an overall complexity algorithm of $O(N^{1/4}\log^{5/4}N)$, using amplitude amplification.

%\break

\section{Scale analysis}\label{sec:anal}

Although an exact analytical description is prohibitive, in this section we will provide some elements of analysis that will convince the reader of the robustness of the numerical results obtained in the previous section. Let's start from the eigenvalues given in Eq.\eqref{ev}, and rescale them using Eq.\eqref{trick}:
\begin{equation}
    E'_{\pm} = \frac{1}{2}g\delta \pm \frac{1}{2}\sqrt{(g\delta)^2+4E^2(1+cg\delta)^2}
\end{equation}
Setting $g(N) = \log N/\pi$ and $c(N) = \sqrt{\dfrac{N \log N}{16 \pi}}$, as in our numerical simulation, we can clearly distinguish two different behaviours for the probability $|a(t)|^2$: 
\begin{itemize}
    \item A strictly linear regime, when $\delta(t)=0$, which appears periodically with a period 
    \begin{equation}
        T_{0} \sim 1/\Delta\lambda \sim \dfrac{1}{E} = O(\sqrt{N \log N})
        \label{eq:T0}
    \end{equation}
    \item A nonlinear regime, when $\delta(t)$ is maximum. In this case, $(1+cg\delta(t)) \sim cg\delta(t) = O(\sqrt{N \log N})$ and consequently the period is constant:
    \begin{equation}
        T_1 \sim 1/\Delta\lambda = O(1).
    \label{T1}
    \end{equation}
\end{itemize}

Now, we have to characterize how the system goes from the first regime to the second one and under which conditions. 
With no loss of generality, and because we are solely interested to the first probability peak behaviour, we focus on the first half period of the dynamics. We propose the following ansatz:  
\begin{equation}
    |a(t)|^2 = \frac{A\pi}{\log N} \sin^2\left(
    \frac{C_0 t}{T_0} + C_1 |a(t)|^2\frac{t}{T_1}
    \right)
\label{self}
\end{equation}
where $A$, $C_0$ and $C_1$ are three real constants. \\

\begin{figure}[h!]
    \centering
    \includegraphics[width = \columnwidth]{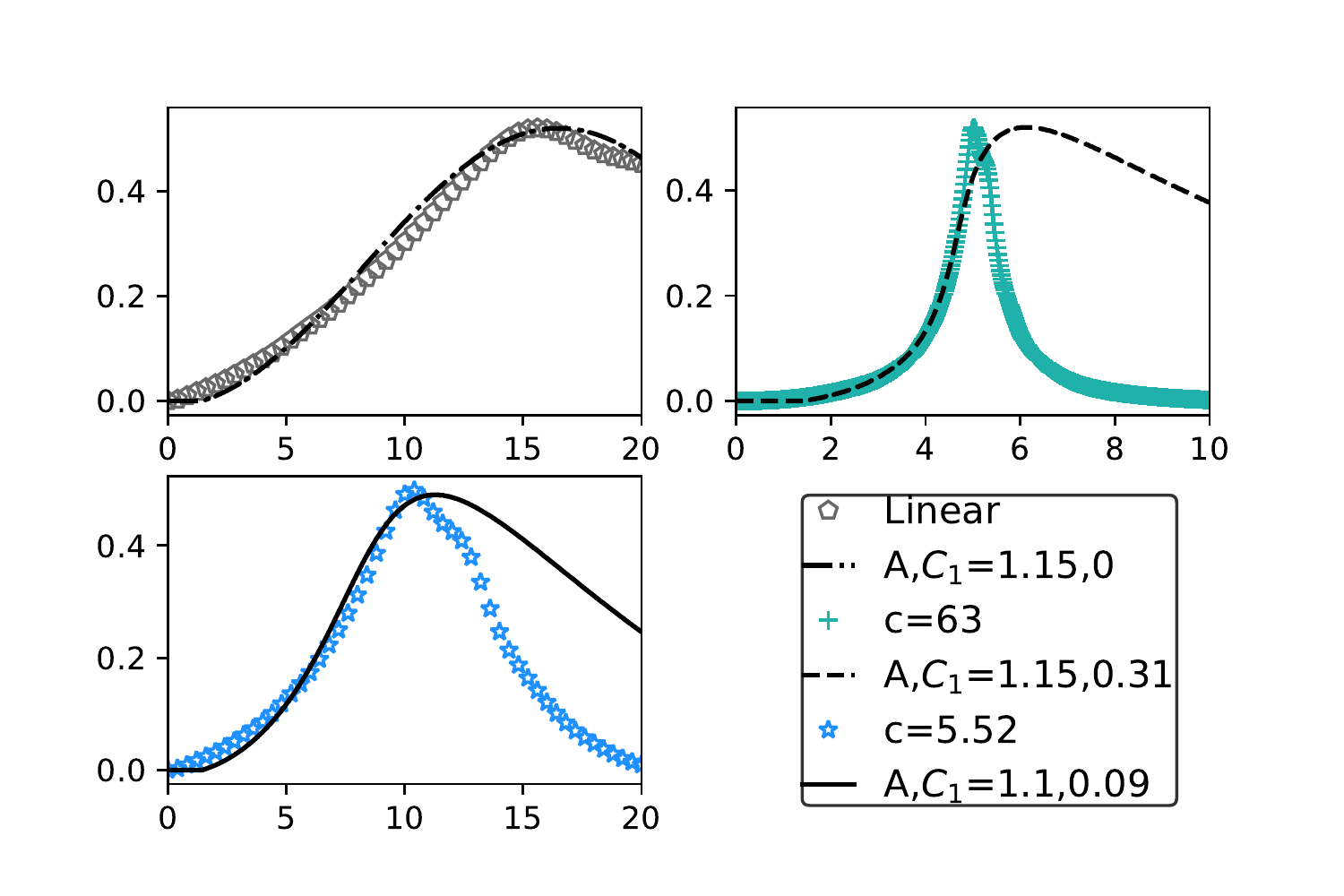}
    \caption{$N=900$ - Probability over $\ket{\Gamma}$ for different values of $c$ and numerical resolution of eq. (\ref{self}).}
    \label{fig:self}
\end{figure}

Notice that the above self-consistent equation describes both regimes for the probability $|a(t)|^2$: indeed , when $\abs{a(t=0)}^2 = 0$, we recover the Eq.\eqref{eqn:overlap}. By choosing $C_0=1$ and $A=1$, and as soon as $\abs{a}^2$ increases, the argument in the \BH{sine} increases as well, speeding-up the dynamics, as observed on Fig. \ref{diff_c}. As $T_1 \ll T_0$, in the limit of large $N$, there exists a critical time $t_s$, at which the Eq.\eqref{self} coincides with $|a(t)|^2 = \frac{A\pi}{\log N} \sin^2\left( C_1 |a(t)|^2\frac{t}{T_1} \right)$. \\ We may argue that $t_s$ scales in the same way than the searching time $T'$ with $N$.
Numerical simulations shows that this is approximately true and it is shown in Fig.\ref{fig:self} for several values of $c$. Along the first half period, the ansatz works surprisingly well. 

Moreover, if we compute the characteristic time $t_s$ at which the system transits from the first behavior with period $T_0$ to the second one with period $T_1$, we derive exactly the same scaling laws, we have for $T$ and $T'$. Indeed, at $t_s$, we have $|\alpha(t_s)|^2 t_s/T_1 \sim t_s/T_0$ and it follows: 
\begin{equation}
    \frac{1}{\log N} \sin^2(t_s/T_0) \sim T_1/T_0.
\end{equation}
In conclusion, by keeping only the first term in the development of $\arcsin{x}$ :
\begin{equation}
    t_s = O(\sqrt{ T_1 T_0 \log N})
\end{equation}
and using Eqs. \eqref{eq:T0} and \eqref{T1}, we finally recover
\begin{equation}
    t_s = O(N^{1/4} \log^{3/4}N),
\end{equation}
which is consistent with the fits realized in the previous section and confirms the time complexity of the nonlinear algorithm that we numerically assessed in Sec. \ref{sec:results}. \\

\section{Discussion}\label{sec:discussion}

\begin{figure}[h!]
    \centering
    \includegraphics[width = \columnwidth]{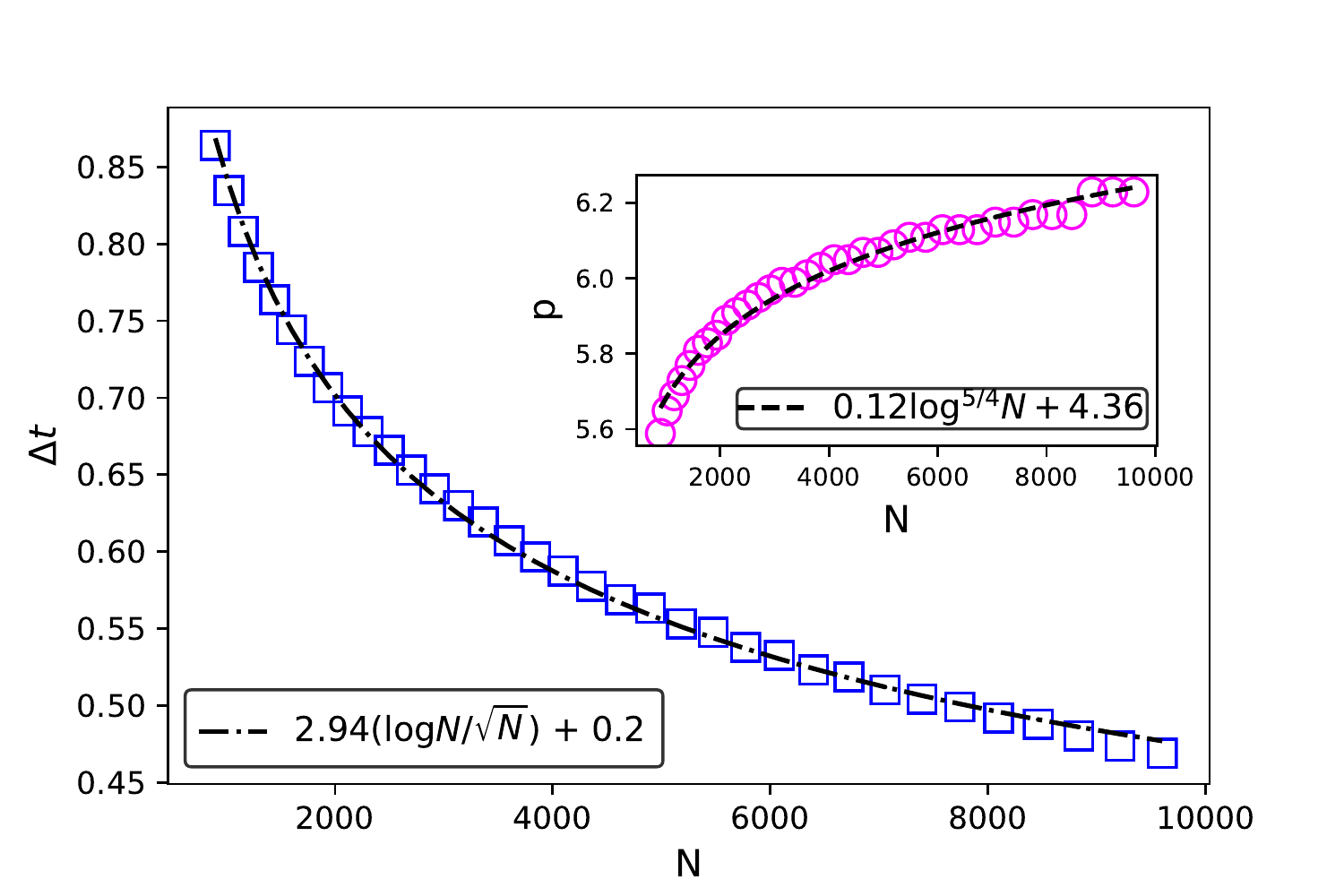}
    \caption{Width of the probability peak - Insight : Searching time}
    \label{width}
\end{figure}

We provide numerical evidence that the nonlinear searching introduced by Wong and Meyer \cite{meyer2013nonlinear} can be extended to the finite 2-dimensional grid, keeping the same computational advantage with respect to the classical algorithm. For this purpose, we have considered a different Hamiltonian, admitting a linear dispersion relation as proved in \cite{Childs_2014}. The numerical simulations showed that the walker finds the marked vertex in $O(N^{1/4} \log^{3/4} N) $ steps with probability $O(1/\log N)$, for an overall complexity of \BH{$O(N^{1/4}\log^{5/4}N)$, using amplitude amplification}. These results are consistent with those proved by Wong and Meyer in \cite{wong2015nonlinear} using a nonlinear Schrödinger equation on a complete graph. However, the optimality of the above algorithm strongly depends on the time measurement precision. As we mentioned earlier, the width of the probability peaks may affect the complexity of the algorithm as already discussed in \cite{wong2015nonlinear}. The narrower the peak, the less efficient the algorithm will be. The optimality of the nonlinear algorithm depends on the choice of $g$ and $c$. For instance, if we had choose both $c$ and $g$ in $O(\sqrt{N})$, we would end up with a runtime of $t_s = O(\log^{5/4} N)$. However, the period $T_1$ of the Eq. \ref{T1} would no longer be a constant, it would rather decrease with $N$ :
\begin{equation}
 T_2 = O\left(\frac{\log N}{\sqrt{N}}\right).
\end{equation}
The peak width at half maximum, shown in Fig.(\ref{width}), is given by : 
\begin{equation}
    \sin^2(\Delta t/2T_2) = \frac{1}{2},
\end{equation}
which yields to
\begin{equation}
    \Delta t \sim T_2.
\end{equation}
Now, considering that $n$ ions are required in an atomic clock to obtain a time precision of $O(1/n)$ \cite{clock}, we have to require an extra $O(1/T_2)$ temporal resource, which dramatically affects the total complexity, making us lose any kind of advantage over the linear algorithm. 

In conclusion, our work presents several advantages with respect to the previous ones: (i) quantum walks are easily implementable in our labs by several physical systems; (ii) graphs having sets of vertices of constant degree, are more natural and pave the way to a $3$-dimensional generalisation. An extension of the above results to other kind of nonlinearities will deserve future investigation.

\section{Acknowledgements} The authors acknowledge inspiring conversations with Thomas Wong. This work has been funded by the Pépinière d’Excellence 2018, AMIDEX fondation, project DiTiQuS and the ID 60609 grant from the John Templeton Foundation, as part of the “The Quantum Information Structure of Spacetime (QISS)” Project.

\clearpage

\bibliographystyle{plain}	
\bibliography{bibli}

\end{document}